\documentclass[11pt,twoside]{article}
\usepackage{asp2010,natbib,url,siunitx,hyperref}
\resetcounters

\newcommand\Halpha{\mbox{H\,$\alpha$}}
\newcommand\CaIIH{\mbox{Ca\,\textsc{ii}\,H}}
\newcommand\CaII{\mbox{Ca\,\textsc{ii}}}
\newcommand\HeI{\mbox{He\,\textsc{i}}}
\newcommand\FeI{\mbox{Fe\,\textsc{i}}}
\DeclareSIUnit\maxwell{Mx}
\DeclareSIUnit\gauss{G}

\begin{document}

\title{Measuring Magnetic Fields in the Solar Atmosphere}
\author{A.~G.~de Wijn
\affil{High Altitude Observatory, National Center for Atmospheric Research, P.O.~Box~3000, Boulder, CO 80303, USA}}

\begin{abstract}
Since the discovery by Hale in the early 1900s that sunspots harbor strong magnetic field, magnetism has become increasingly important in our understanding of processes on the Sun and in the Heliosphere.
Many current and planned instruments are capable of diagnosing magnetic field in the solar atmosphere.  Photospheric magnetometry is now well-established.
However, many challenges remain.
For instance, the diagnosis of magnetic field in the chromosphere and corona is difficult, and interpretation of measurements is harder still.
As a result only very few measurements have been made so far, yet it is clear that if we are to understand the outer solar atmosphere we must study the magnetic field.
I will review the history of solar magnetic field measurements, describe and discuss the three types of magnetometry, and close with an outlook on the future.
\end{abstract}

\section{Introduction}

The importance of magnetic field in astrophysical processes has long been recognised.
Magnetic field is found in all parts of the universe and on all scales: planets, stars, galaxies, accretion disks, etc.
Even in cases where magnetic field was initially deemed too weak to significantly affect plasma dynamics, it has often turned out to be an important factor in the evolution of the system
	\citep[e.g., in accretion disks,][]{1991ApJ...376..214B}.
Nowadays a thriving field of research in astrophysics is centered around the life-cycle of magnetic field.
How is it created?
How does it interact with the plasma it is embedded in?
How is it destroyed?
The diagnosis of magnetic field is now spreading throughout all parts of astrophysics, e.g., Zeeman-Doppler imaging of stars
	\citep{1989A&A...225..456S}
and exoplanets interacting with their host star
	\citep{2003ApJ...597.1092S}.
Indeed, magnetic field has now been inferred to be present or even measured in many astrophysical objects.

\articlefigure[width=0.74\textwidth]{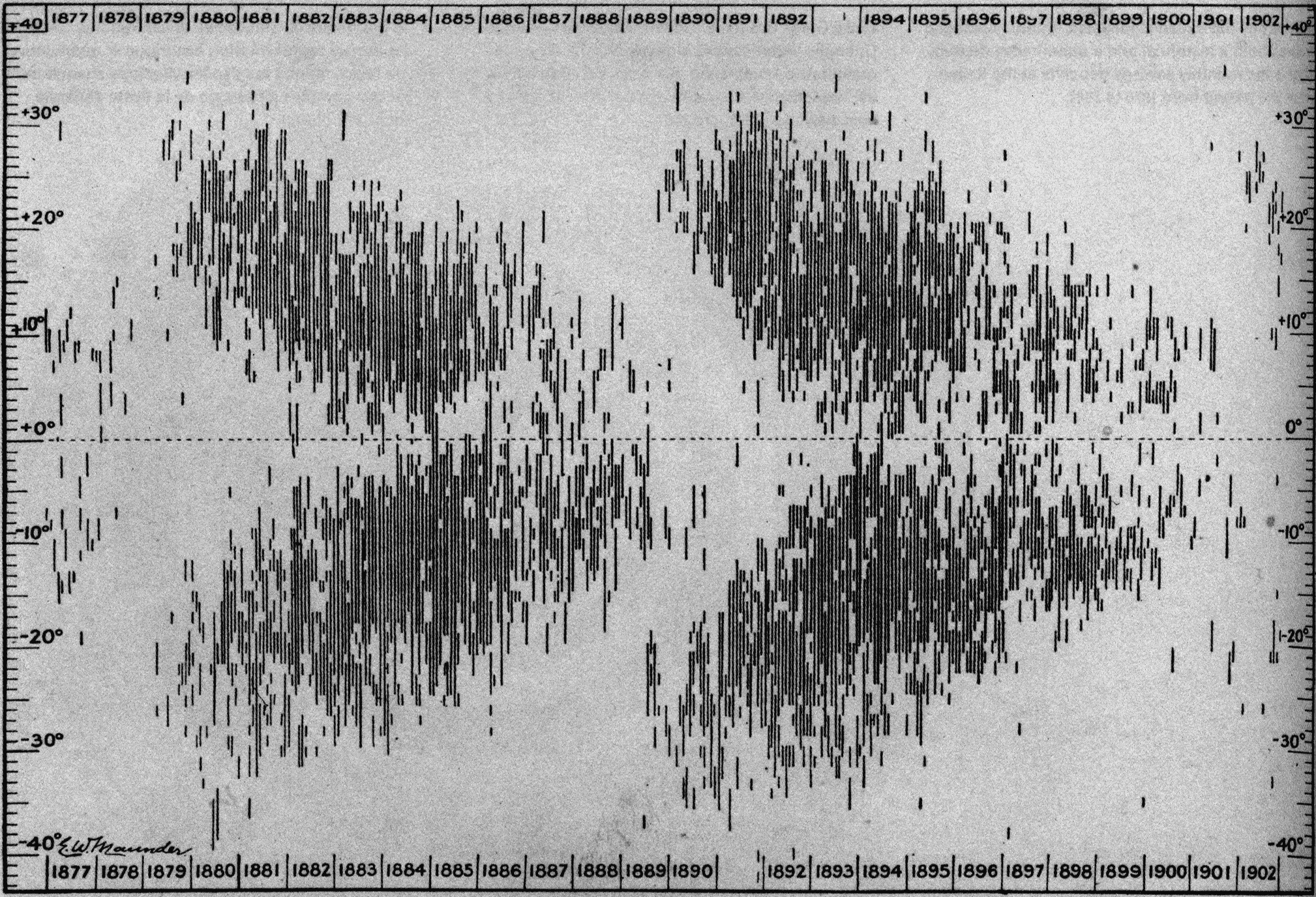}{dewijn:fig:butterfly}{%
The Maunder butterfly diagram showing the latitude variation of active regions over the solar cycle.}

Magnetic fields feature perhaps most prominently in solar physics.
The definite discovery that sunspots have strong magnetic fields was made by
	\cite{1908ApJ....28..315H},
who used a primitive spectro-polarimeter to observe circular polarization in spectral lines resulting from the well-known Zeeman effect
	\citep{Zeeman1896}.
A more detailed publication \citep{Zeeman1897} was reproduced in its entirety that same year in The Astrophysical Journal
	\citep{1897ApJ.....5..332Z}.
These discoveries spurred a century of advancement spectroscopy, polarimetry, instrument development, and research of solar magnetism.
It is worth noting that the first detection of widening of spectral lines in sunspots, an effect now unambiguously attributed to magnetism, was first reported by
	\cite{1866RSPS...15..256L},
more than 30~years prior to the laboratory discovery of the Zeeman effect, and shown by
	\cite{Young1883book}
in a graphic illustration of an observation dated back to 1870.
This subject is covered in detail in the excellent review by
	\cite{1996VA.....40..241D}.

On the cutting edge are instruments like the Helioseismic and Magnetic Imager
	\citep[HMI,][]{2012SoPh..275..229S}
on the Solar Dynamics Observatory
	\citep[SDO,][]{2012SoPh..275....3P},
which provides us with the full-disk vector magnetic field in the solar photosphere several times per hour,
and the Spectro-Polarimeter (SP) of the Solar Optical Telescope
	\citep[SOT,][]{2008SoPh..249..167T}
of Hinode
	\citep{2007SoPh..243....3K},
which combines unparalleled spatial resolution with high sensitivity to make the most refined measurements of solar photospheric magnetic field to date.

\section{Diagnostics}

Broadly speaking there are two types of diagnostics available to us: qualitative and quantitative.
The magnetogram, neither qualitative nor quantitative, is a popular diagnostic that occupies the space in between.
It is often used because of the difficulties associated with performing quantitative diagnostics.

\subsection{Proxy-magnetometry}

Proxy-magnetometry is the technique of determining the locations of magnetic fields through a change in intensity.
Sunspot maps represent the first application of proxy-magnetometry.
Over half a century prior to the discovery of the magnetic nature of sunspots, the 11-year solar cycle was discovered by
	\cite{1844AN.....21..233S},
who noticed periodic behavior after 17~years of observations.
Sunspots also emerge closer to the equator later in the cycle.
Known as Sp\"orer's Law, this feature of the solar magnetic dynamo was first discovered by
	\cite{1858MNRAS..19....1C}
and is beautifully shown in the Maunder butterfly diagram in \autoref{dewijn:fig:butterfly}.
That sunspots harbor strong magnetic fields led to the discovery that the polarity of sunspot pairs is the same in a given solar hemisphere and opposite in the other during the cycle, and that it reverses from one cycle to the next
	\citep{1919ApJ....49..153H}.
Despite more than a century and a half of study, the solar dynamo remains poorly understood
	\citep[for a review see][]{lrsp-2010-1}.

Proxy-magnetometry has also contributed substantially to the understanding of small-scale magnetic fields.
	\cite{1922MNRAS..82..168H}
noticed that if spots do not appear in bipolar form, they are trailed or preceded by regions permeated by small, localized brightenings called `faculae'.
He found that magnetic fields are also present in those regions, and theorized about the existence of `invisible sunspots'.
	\cite{1967SoPh....1..171S} 
first observed small magnetic elements well away from sunspots in the \CaII\ network.
	\cite{1968SoPh....4..142B} 
observed the same features in spectra of plage and noted that they were abundant in the vicinity of sunspots, but much more rare in quiet sun.
Observations by
	\cite{1974SoPh...38...43M} 
had adequate resolution to resolve the network into strings of small ``bright points'' located in intergranular lanes, that change in shape and size on timescales comparable to the lifetime of granules.
The network had previously been associated with kilogauss field by
	\citep{1973SoPh...32...41S}, 
suggesting that many of the observed structures were related.
Direct evidence that faculae, gaps, and bright points were all manifestations of the same phenomenon was eventually provided by high-resolution observations, simultaneous in multiple wavelengths, of both disk and limb targets in a study by
	\cite{1981SoPh...69....9W}. 
The bright points form a dense pattern in plage and active network, while outside of active regions, they clump in patches that partially outline supergranular cells.

The relative ease of observing magnetic fields via proxy-magnetometry resulted in extensive studies of bright points in the photospheric network.
Imaging in wide-band \CaIIH\ and K or \Halpha\ were the diagnostics of choice, until
	\cite{1984SoPh...94...33M} 
switched to using the Fraunhofer G~band around \SI{430.8}{\nano\meter} in order to reduce the effects of chromatism in their telescope.
The blue wavelength of the G~band allows for both high resolution when compared to \Halpha\ as well as good sensitivity of film and silicon detectors when compared to \CaIIH\ and K.
It is still widely used as one of the principal diagnostics for proxy-magnetometry
	(\citealp[e.g.,][]{2005A&A...435..327R,2012ApJ...752...48C};
	and references in \citealp[Sect.~5 of][]{2009SSRv..144..275D}).

The suggestion by 
	\cite{1981SoPh...70..207S} 
that facular brightness enhancement is the result of radiation escape from hot granular walls was confirmed by 3D MHD simulation by
	\cite{2004ApJ...607L..59K} and 
	\cite{2004ApJ...610L.137C}. 
They modeled flux tubes, then ``observed'' them as if close to the limb using radiative-transfer codes to calculate the emergent intensity.
Partial evacuation of the flux tube allows the observer to look deeper into the hot granular wall than would be possible if the flux tube were absent.

\articlefigure[width=0.75\textwidth]{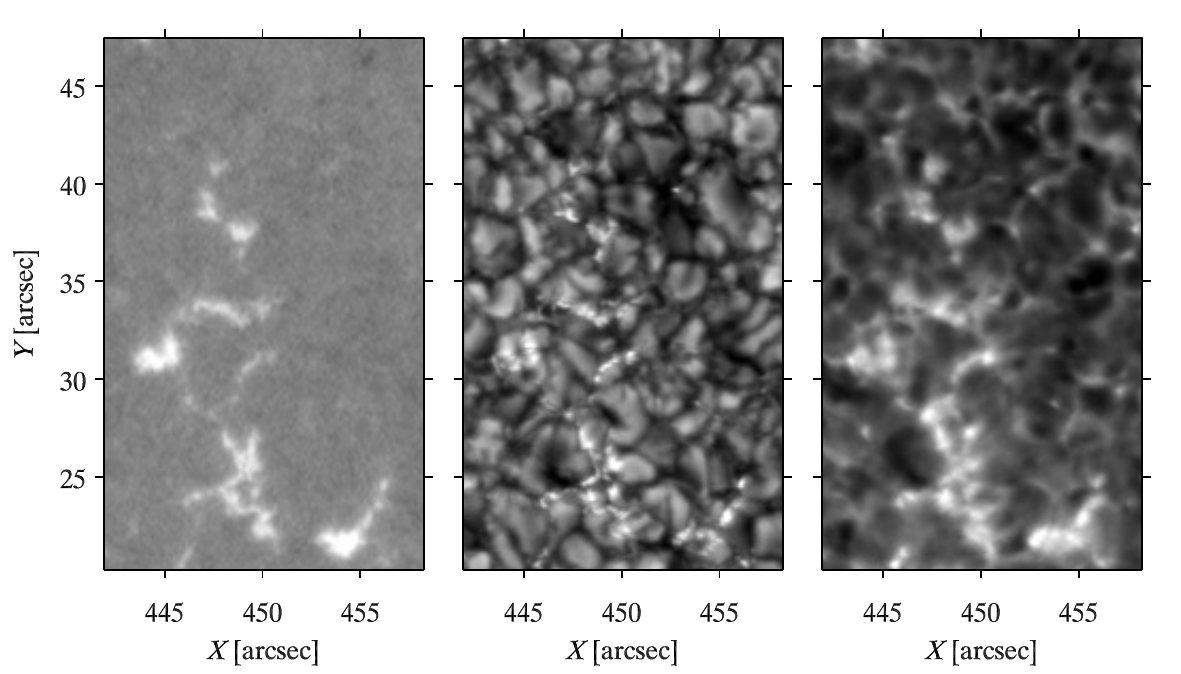}{dewijn:fig:proxysample}{%
Sample network region.
Left: calibrated \FeI\ \SI{630.2}{\nano\meter} line-of-sight magnetogram, scaled between \num{-1e3} (black) and \SI{e3}{\maxwell\per\square\centi\meter} (white).
Middle: G-band intensity.
Right: \CaIIH\ intensity.
Bright points in the G~band and in the \CaIIH\ line images correlate well with positions of concentrated field.
The network consists of strings of several adjacent bright points, located in the intergranular lanes.
Bright points in the \CaIIH\ image appear more extended and fuzzy than in the G-band image due to expansion of the flux tube with height.
These images were taken with Hinode/SOT on March~30, 2007, around 00:24:30~UT.
The coordinates indicate distance from sun center, so that $\mu\approx0.88$.
From \cite{2009SSRv..144..275D}.}

\subsection{Magnetograms}

The next step up from proxy-magnetometry is the ``magnetogram'', occupying the space between a qualitative diagnostic like G-band imagery and a quantitative diagnostic such as spectro-polarimetric inversions.
Magnetograms are made using a narrow-band filter combined with a simple polarimeter.
By tuning the filter to the wing of a magnetically sensitive line and observing circular polarization, a measure of the line-of-sight magnetic field is found.
While the measurement is not truly quantitative, it can be calibrated with some accuracy to provide magnetic flux density, i.e., the product of the filling factor and strength of the magnetic field.
Because magnetographs are only sensitive to circular polarization, they only provide a diagnostic to the line-of-sight component of the magnetic field.
\autoref{dewijn:fig:proxysample} displays an example of a magnetogram together with proxy-magnetometry using imaging in the G~band and in the \CaIIH\ line.

Perhaps the best known example of a magnetograph is the Michelson Doppler Imager
	\citep[MDI,][]{1995SoPh..162..129S}
on the Solar and Heliospheric Observatory
	\citep[SOHO,][]{1995SoPh..162....1D}.
The uninterrupted, long-duration, and seeing-free MDI observations led to a huge increase in the understanding of, e.g., the formation of the photospheric magnetic network
	\citep{1997ApJ...487..424S} and
dispersal of magnetic flux
	\citep{1999ApJ...511..932H}.

\subsection{Quantitative Measurement}

The advent of the CCD camera for the first time allowed direct digitization of observations and consequent analysis by computers, and as computers became more powerful it became feasible to compute spectral lines based on a polarized radiative transfer calculation.
This development led to the process now commonly referred to as ``inversions''.
It is described in detail in \autoref{dewijn:sec:inversions}.
Simply put, from an observation of a spectral line one can derive the plasma parameters in the region of the solar atmosphere where that line formed.

The instrument of choice for these kinds of measurements is known as a spectro-polarimeter: the combination of a spectrograph and a full-Stokes polarimeter.
The typical layout of such an instrument consists of the telescope feed optics, followed by a polarimetric modulator and a tradition spectrograph.
A polarimetric modulator is a device that alters the polarized light entering the telescope in a known way.
Because detectors are sensitive to intensity only, an `analyzer', typically a linear polarizer, is then used to imprint polarization information into intensity.
Out of at least four independent measurements one can reconstruct the Stokes vector consisting of intensity (I), linear polarization (Q and U), and circular polarization (V).
An example of a spectro-polarimetric observation, taken with the Hinode/SOT Spectro-Polarimeter, is shown in \autoref{dewijn:fig:sp}.
The spectro-polarimeter of
	\cite{1908ApJ....28..315H}
could only modulate Stokes V, and as a result it was only capable of line-of-sight diagnostics.
While modern spectro-polarimeters are significantly more sophisticated than that instrument, the principle of operation remains the same.

The first data to ever be processed with an inversion technique was taken with the Stokes~I spectro-polarimeter developed at the High Altitude Observatory (HAO) and deployed at the Dunn Solar Telescope at Sac Peak.
The success of Stokes~I and its successor Stokes~II led to the development of the Advanced Stokes Polarimeter
	\citep[ASP,][]{1992SPIE.1746...22E}
and many other instruments in the early 1990s, giving scientists a new window on magnetism in the solar photosphere
	\citep[e.g.,][]{1987ApJ...318..930L,1993ApJ...418..928L}.
As an example, the SOT Spectro-Polarimeter on Hinode is a derivative of ASP.

Narrow-band imaging instruments have come a long way since the first birefringent filters were built by
	\cite{1944AnAp....7...31L}.
Some instruments have sufficient spectral resolution that they can be used as true spectro-polarimeters.
Fabry-Perot interferometers
	\citep{1902ApJ....15...73F}
in particular have been very successful
	\citep[e.g.,][]{2008ApJ...689L..69S,2009ApJ...700L.145V,2010A&A...520A.115B,2010ApJ...710.1486J}.
The same computational techniques used with data from spectrographs for the inversion of plasma parameters can be applied to data from these instruments.
The benefit of an imaging spectro-polarimeter is that it samples a 2D region, and usually takes only a few seconds to make a spectral scan.
A slit-scanning spectro-polarimeter must raster to build up an image, a process that can take many minutes.
The slow scanning is traded against high spectral resolution and high sensitivity.

\section{Inversions}\label{dewijn:sec:inversions}

\autoref{dewijn:fig:inversions} shows a schematic overview of the process of spectro-polarimetric inversions.
While this process is called an inversion, it really involves solving the forward problem of synthesizing an observation based on an atmospheric model and external fields, and comparing that synthesis to the actual observations.
The fields and the model are then adjusted until a match is found.

\articlefigure[width=0.75\textwidth]{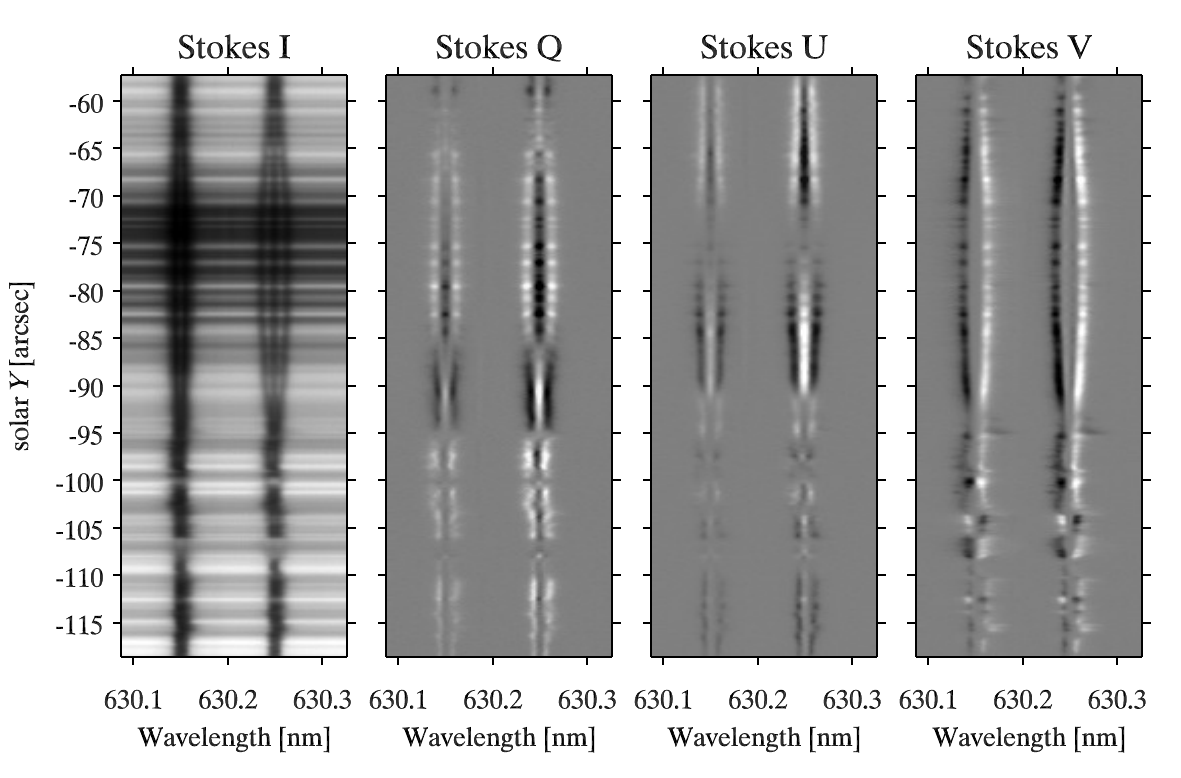}{dewijn:fig:sp}{Example of spectro-polarimetric data taken by the Hinode/SOT Spectro-Polarimeter instrument on Dec 12 2006 at 11:50:04~UT.
The $X$ position of the slit was \SI{242.5}{\arcsec} W of disk center.
From left to right: intensity (Stokes I), linear polarization (Stokes Q and U), and circular polarization (Stokes V).
Around $Y=-75\si{\arcsec}$ the spectrograph slit crosses a sunspot.
The Zeeman effect causes the intensity profiles to be broadened in the \FeI\ \SI{630.15}{\nano\meter} line and separated into three components in the \FeI\ \SI{630.25}{\nano\meter}.
The blue and red components are circularly polarized in opposite directions, as can be seen in the Stokes V image.}

An atmospheric model is chosen and a guess is made of the external parameters.
The radiation output can be calculated through radiative transfer if the atomic excitation is known.
The atomic excitation, however, relies on a statistical equilibrium calculation that depends on the radiation field.
This loop must be iterated until convergence is reached.
Then, the radiation output can be compared with the observations.
If the quality of the match is found to be insufficient, a new set of parameters must be evaluated, until a set is found that provides a good match.

Many photospheric lines are formed in local thermal equilibrium (LTE), and statistical equilibrium does not need to be considered.
Atomic excitation can be calculated directly from the plasma parameters.
The calculation of the radiation output is fast because the consistency loop is broken.
The simplest so-called `Milne-Eddington' atmosphere, in which the continuum and line source functions are the same and radiative transfer is easy to compute, is commonly used.
Not surprisingly a method by
	\cite{1977SoPh...55...47A}
that used this approximation was the basis for the first applications of inversions of spectro-polarimetric observations
	\citep{1982ApJS...49..293L,1987ApJ...322..473S}.
It remains the workhorse for inversions of photospheric vector magnetic field
	\citep{2011SoPh..273..267B}.
More sophisticated LTE atmospheres may, e.g., derive height dependent information.
They have more parameters than the Milne-Eddington atmosphere and are consequently computationally more expensive.

For lines formed in non-LTE (e.g., those that sample the chromosphere), these approaches are not sufficient.
The full process has to be evaluated, requiring knowledge of the radiation field as well as a time-consuming iteration of the loop to ensure consistency.
However, in some cases it can be assumed that the single-scattering approximation is valid.
If so, the statistical equilibrium calculation does not require the radiation output, breaking the loop and making the problem tractable.

The codes that perform these calculations must find a global minimum in multi-dimensional parameter space that usually has many local minima.
For the simpler atmospheric models, i.e., Milne-Eddington and those using the assumption of LTE, the code will usually start the optimization using a Monte-Carlo method or genetic optimization in order to find the global minimum.
As the solution improves, the code will switch to a derivative-based method to finalize the optimization.
An example of an inversion is shown in \autoref{dewijn:fig:inversion}.

\articlefigure[width=0.75\textwidth]{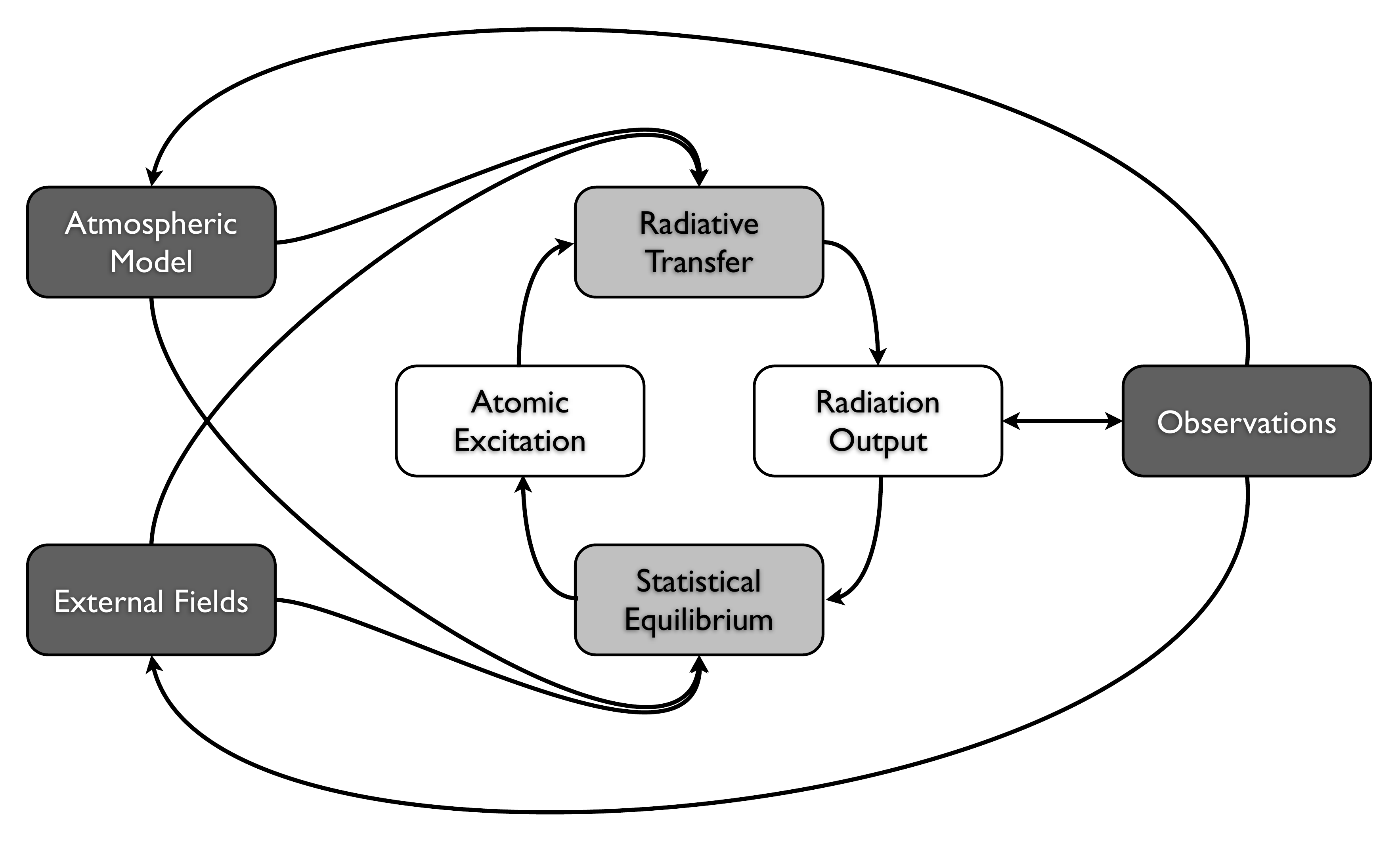}{dewijn:fig:inversions}{Schematic overview of the process of spectro-polarimetric data inversion.
In dark grey are inputs to the process: the observations that are to be inverted, the atmospheric model chosen by the scientist, and the external fields.
In white are outputs generated by the radiative transfer and statistical equilibrium calculations (light grey).}

For the non-LTE case, even under the assumption of single-scattering, the calculations are typically too costly to use a brute-force approach to find the global minimum, and a smarter approach is called for.
Codes have been developed that find the solution through a technique known as principle component analysis.
A large database of lines is computed for many different parameter sets.
This database is then decomposed into a basis of components called principle components.
Once the database has been computed and the basis has been established, an observed line can be quickly approximated by a linear combination of elements of this basis, thus yielding an inversion of the observation.
This method has been applied successfully to spectro-polarimetric observations of chromospheric structures such as solar prominences
	\citep{2005ApJ...622.1265C},
and active region filaments
	\citep{2009A&A...501.1113K}.

Regardless of the inversion strategy, the nature of the data leaves us with ambiguities.
The \SI{180}{\degree} ambiguity in the determination of the azimuth is well-known.
A number of manual and automatic processes have been developed that attempt to remove this ambiguity by making assumptions about the magnetic field configuration that typically involves minimizing some quantity that depends on the azimuth of the magnetic field, e.g., the vertical gradient of the magnetic pressure.
However, it is known that at least in some areas of the Sun these quantities are not in a minimum state, and so the resolution cannot be expected to yield the correct result.
A good overview of this problem and the performance of several algorithms is given in 
	\cite{2006SoPh..237..267M}.

\articlefigure[width=0.75\textwidth]{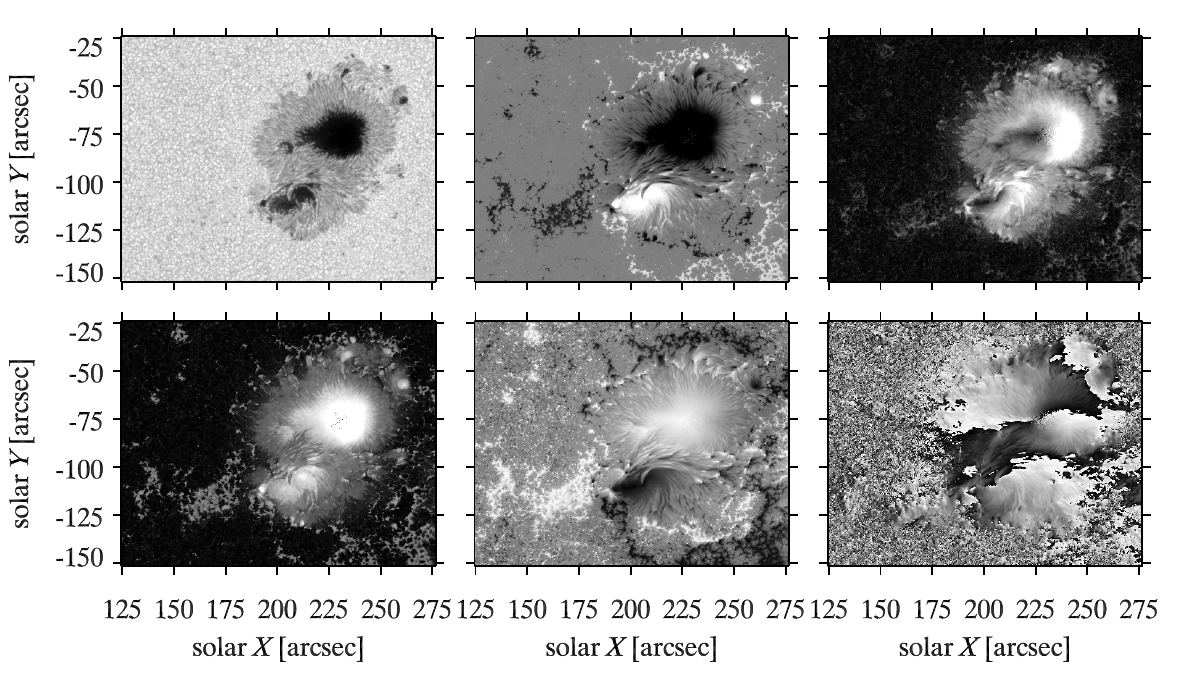}{dewijn:fig:inversion}{Example of an inversion of spectro-polarimetric data taken by the Hinode/SOT Spectro-Polarimeter instrument on Dec 12 2006 from 10:10:08~UT to 13:11:18~UT.
The data shown in \autoref{dewijn:fig:sp} was taken from this map.
Clockwise from the top left (black to white scaling): continuum intensity (arbitrary), line-of-sight component of the magnetic field (\SIrange[range-units = single]{-2}{2}{\kilo\gauss}), transverse component (\SIrange[range-units = single]{0}{2}{\kilo\gauss}), azimuth (\SIrange{0}{180}{\degree}, i.e., with \SI{180}{\degree} ambiguity), inclination angle (\SIrange{0}{180}{\degree}), and field strength (\SIrange[range-units = single]{0}{3}{\kilo\gauss}).}

\section{Future Developments}

Many instruments now regularly provide us with a measure of the vector magnetic field in the solar photosphere.
The vector magnetic field in the photosphere is determined a few times per hour from full-disk SDO/HMI data.
While such measurements have greatly improved our understanding of the solar atmosphere and Sun-Earth interactions, it is clear that they are not sufficient to resolve a number of long-standing problems, such as coronal heating, the origin of the solar wind, and acceleration of coronal mass ejections.
The key to all these unresolved issues lies with the magnetic field in the chromosphere and the corona.

One can attempt to extrapolate the photospheric field measurements into the chromosphere and corona.
Unfortunately, this process is difficult
\citep[see, e.g.,][]{2009ApJ...696.1780D}.
Potential field extrapolations are clearly incorrect.
More sophisticated methods have been applied with some success, but rely on ``preprocessing'' of the photospheric magnetic field in such a way that the extrapolated field approximates the morphological structure observed in a diagnostic of the chromosphere (typically \Halpha\ or IR \CaII).
A direct measure of the chromospheric vector magnetic and velocity field over the full solar disk is highly desirable to forgo this step, as it provides a true force-free boundary at the base of the heliosphere.

The first successful measurements of coronal magnetic field have been made by
	\cite{2004ApJ...613L.177L}
using a fiber-fed integral-field spectro-polarimeter and by
	\cite{2008SoPh..247..411T}
using the Coronal Multi-channel Polarimeter (CoMP).
All coronal field measurements by necessity employ a coronagraph that blocks out the light from the solar disk.
Consequently it is only possible to determine the magnetic field integrated with some weighing along the line of sight in the plane of the sky.

A number of instruments have been or are currently being developed with the intent of measuring magnetic field in the chromosphere, focusing on four lines that are thought to be the most promising: the IR \CaII\ triplet and \HeI\ \SI{1083.0}{\nano\meter}.
The latter is easier to interpret and has been used with some success to measure magnetic field in prominences, active region filaments, and spicules
	\citep{2005ApJ...622.1265C,2009A&A...501.1113K,2010ApJ...708.1579C}.
However, in quiet sun the line can be three orders of magnitude weaker than in active regions.
The IR \CaII\ triplet at \SIlist{849.8;854.2;866.2}{\nano\meter} is strong everywhere, but suffers in other aspects: the radiative transfer calculations required to model it are difficult, it forms over a broad height range, the lines are spectrally far apart, and \SIlist{849.8;866.2}{\nano\meter} have blends.
However, these four lines are still the most promising ones in the visible and near-infrared spectrum.

The Coronal Solar Magnetism Observatory (COSMO) has been proposed as a facility for synoptic measurements of the chromospheric and coronal magnetic field.
It is to contain a suite of three instruments: a white-light coronagraph for coronal density measurements (under construction), an instrument for chromospheric magnetic field diagnostics (under development), and a large coronagraph for coronal field measurements (in preliminary design phase).

It is also evident from theoretical considerations and simulation alike
	\citep[e.g.,][]{2009Sci...325..171R}
that high-resolution magnetometry is another frontier that must be explored.
Here, we have reached the limit of present-day telescopes.
In the near future results are expected from the \SI{1.6}{\meter} New Solar Telescope at the Big Bear Solar Observatory and the \SI{1.5}{\meter} GREGOR telescope on Tenerife, Spain.
Under construction and development, respectively, are the Advanced Technology Solar Telescope and the European Solar Telescope.
Both will have \SI{4}{\meter} apertures, are designed with polarimetry in mind, and will be outfit with imaging and slit-scanning spectro-polarimeters.

\acknowledgements The National Center for Atmospheric Research is sponsored by the National Science Foundation.

\end{document}